\documentclass[aps,prb,superscriptaddress,twocolumn]{revtex4-1}

\usepackage{graphicx}
\usepackage{bm}
\usepackage{amsmath}

\newcommand{\beq}{\begin{equation}}
\newcommand{\eeq}{\end{equation}}
\newcommand{\beqa}{\begin{eqnarray}}
\newcommand{\eeqa}{\end{eqnarray}}

\newcommand{\w}{\omega}

\newcommand{\ket}[1]{\left| #1 \right\rangle}

\begin{document}
\title{Ground state and dynamics of the biased dissipative two-state system: Beyond variational polaron theory}
\author{Ahsan~Nazir}\email{a.nazir@imperial.ac.uk}
\affiliation{Blackett Laboratory, Imperial College London, London SW7 2AZ, United Kingdom}
\author{Dara P. S. McCutcheon}
\affiliation{Blackett Laboratory, Imperial College London, London SW7 2AZ, United Kingdom}
\affiliation{Department of Physics and Astronomy, University College London, Gower Street, London WC1E 6BT, United Kingdom}
\affiliation{London Centre for Nanotechnology, University College London, United Kingdom}
\author{Alex W. Chin}
\affiliation{Theory of Condensed Matter Group, University of Cambridge, J J Thomson Avenue, Cambridge, CB3 0HE, United Kingdom}

\date{\today}

\begin{abstract}

We propose a ground-state ansatz for the Ohmic spin-boson model that improves upon the variational treatment of Silbey and Harris for biased systems in the scaling limit. In particular, it correctly captures the smooth crossover behaviour expected for the ground-state magnetisation when moving between the delocalised and localised regimes of the model, a feature that the variational treatment is unable to properly reproduce, while it also provides a lower ground-state energy estimate in the crossover region. We further demonstrate the validity of our intuitive 
ground-state by showing that it leads to predictions 
in excellent agreement with those derived from a non-perturbative Bethe-ansatz technique. 
Finally, recasting our ansatz in the form of a generalised polaron transformation, we are able to explore the dissipative two-state  
dynamics beyond weak system-environment coupling within 
an efficient time-local master equation formalism.

\end{abstract}

\maketitle

\section{Introduction}
The thermodynamics and non-equilibrium dynamics of quantum systems in contact with environmental degrees of freedom is a topic of primary importance in  
physics and chemistry,~\cite{leggett87,weissbook,maykuhnbook,nitzanbook} and is also becoming increasingly relevant in biology for systems in which quantum effects may play an important role.~\cite{engel07,calhoun09,collini09,collini09sc,collini10,panitchayangkoona10} In typical experiments, it is impossible to observe the degrees of freedom of the environment. The unmeasured correlations which build up between the system and environment then lead to an effectively irreversible, non-unitary dynamics of the reduced state of the quantum system. This 
is often classified into two fundamental processes, the intuitive relaxation of the system to thermal equilibrium, caused by energy exchange with the environment, and the destruction of quantum mechanical coherence between classical system states, 
known as decoherence.

In situations where the environment is weakly coupled to the system, methods such as Redfield or Lindblad theory can be applied,~\cite{b+p} and the dynamics of the reduced state can be described using simple time-local master equations. However, when the coupling is strong, or if the decay of environmental correlations is slow, then the Born-Markov approximation on which these techniques are based will fail, and a more sophisticated treatment of system-bath correlations and bath-memory effects 
is required. The development of methods to treat such cases has recently been necessitated by interest in a wide range of quantum systems in which the environmental interactions and 
dynamics are non-trivial, leading to complex reduced system state dynamics which are intermediate between semi-classical energy relaxation and purely quantum coherent (wavelike) motion. Important examples include superconducting and spin quantum dot qubits for quantum computation~\cite{kit2,coish2004hyperfine,bellomo2007non} and the recently discovered long-lasting wavelike motion of excitons in photosynthetic pigment-protein complexes.~\cite{engel07,calhoun09,collini09,collini09sc,collini10,panitchayangkoona10} 

The extensively-studied spin-boson 
model has established itself as perhaps the most important system for developing theoretical concepts and numerical techniques through which we may understand the microscopic behaviour of open quantum systems in all of the regimes mentioned above. The model consists of a quantum two-level system (TLS) that is coupled to an environment made up of a bath (continuum) of harmonic oscillators.  The environment and its couplings to the system are characterised by a spectral function $J(\omega)$, to be defined later. In many important applications, the spectral function behaves as a power-law at low frequencies $J(\omega)\propto \omega^{s}$, and this is often used to classify system-environment coupling types into three distinct groups: sub-Ohmic ($s<1$), Ohmic ($s=1$) and super-Ohmic ($s>1$). Despite its apparent simplicity, the dynamics of the spin-boson problem cannot be solved exactly, and the extremely interesting, and as yet unexhausted, physics in the 
model continues to drive research into its properties and 
the potential implications for quantum devices and bio-complexes.~\cite{brumer,gilmore06,bellomo2007non,fleming2011quantum,huelga2011quantum, shi2011quantum,lei2011decoherence,zhang2011deterministic}

The super-Ohmic case is arguably the simplest to describe and applies to a wide variety of physical systems, such as in the electron-phonon interactions of impurities in solids~\cite{weissbook,wurger98} and quantum dots,~\cite{ramsay10,ramsay10_2,krummheuer02,calarco03} and in atom-photon interactions.~\cite{b+p} However, even in this case a crossover from coherent to incoherent dynamical behaviour is expected as the environmental influence becomes strong,~\cite{wurger98,nazir09,mccutcheon11} and the simplest weak system-bath coupling treatments will then fail. The sub-Ohmic case leads to strongly non-Markovian dynamics and also contains a quantum phase transition in the ground state.~\cite{spohn85,kehrein, vojta2005quantum, winter09,Alvermann09,LeHur07, chin11, Zhang10,florens2011dissipative,zhao11} Sophisticated numerical methods, capable of treating the many-body correlations which drive these phenomena,~\cite{Alvermann09,winter09,wang2008coherent,bullarev08,Anders07,Prior10,chinbook11, nalbach2010ultraslow}  are often required to look at this case, and several artificial systems have been proposed in which these effects could be observed.~\cite{porras2008mesoscopic,recati2005atomic} The Ohmic case lies on the boundary of the non-Markovian, 
many-body physics of the sub-Ohmic regime and the effectively 
Markovian (though not necessarily weak-coupling~\cite{wurger98,nazir09,mccutcheon11}) physics which emerges in the super-Ohmic case. 
Its importance lies in the combination of its non-trivial dynamical properties, which encapsulates a number of different phases, and the many physical realisations of this type of environment.~\cite{weissbook,b+p,leggett87} A particularly 
topical example is found in exciton transport in pigment-protein complexes, where the (Ohmic) over-damped Brownian oscillator model is widely employed as a simplified way to model solvent environments and protein fluctuations.~\cite{ishizaki2010quantum,gilmore06} 

To treat the dynamics of the spin-boson model, various advanced numerical and analytical methods have been proposed and applied.~\cite{leggett87,weissbook} Amongst these, a popular and powerful approach has been the polaron transformation technique,~\cite{nitzanbook,mahan,wurger98,bog2000polaron} which uses a unitary transformation of the spin-boson Hamiltonian to treat part of the system-bath coupling non-perturbatively, and then employs perturbation theory in the residual system-bath couplings to derive a dissipative master equation for the TLS.~\cite{silbey1980general,nazir09,mccutcheon10_2,mccutcheon11,jang08,jang09,kolli11,jang11} The perturbative treatment of residual interactions essentially drives thermalisation amongst the renormalised states of the non-perturbative part of the transformed Hamiltonian, and at low temperatures it is important that the transformation correctly captures the ground state of the system embedded in the environment. Unfortunately, the naive application of standard polaron theory to Ohmic environments \emph{fails}, as all coherent matrix elements are renormalised to zero in the non-perturbative part of the Hamiltonian for all coupling strengths. This leads to purely incoherent dynamics in a time-local master equation approach,~\cite{aslangul85} even though more sophisticated numerical and analytical techniques, as well as common sense, show that damped coherent dynamics take place at sufficiently weak coupling.~\footnote{In fact, using a combination of the polaron transformation and a perturbative time {\emph {non-local}} master equation does lead to damped coherent dynamics at weak coupling, and has been shown to be equivalent to the non-interacting blip approximation.~\cite{leggett87,PhysRevE54R3086,aslangul86} The ground state of the biased two-level system is still incorrectly represented by such a theory, however.}
An improvement on the standard polaron theory is the variational polaron treatment of Silbey and Harris (SH),~\cite{silbey84,harris85,silbey89} in which the zeroth-order Hamiltonian may describe coherent dynamics for weak coupling. However, for biased TLSs this theory predicts an unphysical, discontinuous crossover to incoherent dynamics at a finite coupling strength.~\cite{silbey89} 

The failure of these approaches lies in the improper choice of zeroth-order Hamiltonian, which at low temperatures results in the system relaxing to a ground state which is qualitatively different from the true ground state. In this paper we propose a new ground state ansatz for the biased Ohmic spin-boson model, and show that it predicts results in excellent agreement with non-perturbative treatments based on the Bethe-ansatz for this problem.~\cite{lehur08} As well as being conceptually simpler than these often costly techniques, our ansatz is itself based on a generalised polaron-type transformation that permits the nonequilibrium dynamics of the TLS to be explored using many of the advances recently made in polaron theory.~\cite{jang08,jang09,jang11,mccutcheon10_2,kolli11,mccutcheon11_2,mccutcheon11_3,zimanyi12} 

The paper is set out as follows. In Section~\ref{sec:Model} we introduce the spin-boson model and formulate the problem. 
We then present our microscopic ansatz for the ground state and compare it to the SH theory and the exact Bethe-ansatz solutions of an equivalent theory. 
In Section \ref{unitary} we demonstrate how the ansatz can be recast as a unitary transformation of the original problem, and derive the effective Hamiltonian with which we then compute the TLS dynamics.  
These results and the comparison with the other theories are then discussed. In Section~\ref{sec:other} we briefly comment on the application of our ansatz to non-Ohmic spectral densities, before summarising in Section~\ref{sec:Conclusions}.      

\section{Model}
\label{sec:Model}

The spin-boson Hamiltonian can be written as
\begin{equation}\label{Hspinboson}
H=\frac{\epsilon}{2}\sigma_z-\frac{\Delta}{2}\sigma_x+\sum_{\bf k}\omega_{\bf k}b_{\bf k}^{\dagger}b_{\bf k}+\sigma_z\sum_{\bf k}g_{\bf k}(b_{\bf k}^{\dagger}+b_{\bf k}),
\end{equation}
describing a TLS characterised by a bias $\epsilon$ and tunneling amplitude $\Delta$ between basis states $|1\rangle$ and $|0\rangle$, 
linearly coupled to a harmonic oscillator bath of mode 
frequencies $\omega_{\bf k}$, with 
strengths $g_{\bf k}$ (assumed real).  The standard Pauli operators used above are $\sigma_z=|0\rangle\langle0|-|1\rangle\langle1|$ and $\sigma_{x}=|1\rangle\langle 0| +|0\rangle\langle 1|$. As has been well established by previous studies,~\cite{leggett87,weissbook} for Gaussian initial states the effects of the environment on the reduced state of the TLS in this model are completely determined by the spectral function $J(\omega)$, defined as $J(\omega)=\sum_{\bf k}g_{\bf k}^{2}\delta (\omega_{\bf k}-\omega)$. In this paper we will only consider the Ohmic spectral density, which we parameterise as 
\begin{equation}\label{spectraldensity}
J(\omega)=(\alpha/2)\omega\theta(\omega_{c}-\omega),
\end{equation}
with $\alpha$ being a dimensionless measure of the system-environment coupling strength, while $\theta(x)$ is the Heaviside step function that provides a cut-off to the spectral function at a typical frequency of $\omega_{c}$. 



\subsection{Ground state ansatz}
\label{sec:ansatz}

For computing non-perturbative dynamics in a standard polaron-type theory, it is essential that the zeroth-order Hamiltonian possesses a ground state which is a very good approximation to the true ground state. We shall first construct such a ground state ansatz, and present an equivalent generalised polaron transformation 
in Section \ref{unitary}. The ansatz is composed from the basis $\{|0\rangle\prod_{\bf k}D(\alpha_{{\bf k},0})|\rm{vac}\rangle,|1\rangle\prod_{\bf k}D(\alpha_{{\bf k},1})|\rm{vac}\rangle\}$, where $|\rm{vac}\rangle$ denotes the vacuum state of the bosonic bath and 
$D(\alpha_{\bf k})=\mathrm{exp}[\alpha_{\bf k}(b_{\bf k}^{\dagger}-b_{\bf k})]$ are bosonic displacement operators.~\cite{glauber63} 

In the new (restricted) basis, we may write the spin-boson Hamiltonian as
\begin{equation}
\label{Hspinbosonnewbasis}
\tilde{H}=\frac{(\epsilon+A_0-A_1)}{2}\tilde{\sigma}_z-\frac{\Delta_r}{2}\tilde{\sigma}_x+\frac{(A_0+A_1)}{2},
\end{equation}
where $\tilde{\sigma}_z=|0\rangle\prod_{\bf k}D(\alpha_{{\bf k},0})|\rm{vac}\rangle\langle\rm{vac}|\prod_{\bf k}D(-\alpha_{{\bf k},0})\langle0|-|1\rangle\prod_{\bf k}D(\alpha_{{\bf k},1})|\rm{vac}\rangle\langle\rm{vac}|\prod_{\bf k}D(-\alpha_{{\bf k},1})\langle1|$, $A_{0}=\sum_{\bf k}\alpha_{{\bf k},0}(\omega_{\bf k}\alpha_{{\bf k},0}+2g_{\bf k})$, $A_{1}=\sum_{\bf k}\alpha_{{\bf k},1}(\omega_{\bf k}\alpha_{{\bf k},1}-2g_{\bf k})$, and the tunneling term has been renormalised such that
\begin{eqnarray}\label{deltardefn}
\Delta_r&=&\Delta\langle{\rm vac}|\prod_{\bf k}D(\pm(\alpha_{{\bf k},0}-\alpha_{{\bf k},1}))|\rm{vac}\rangle\nonumber\\
&=&\Delta \exp{\left[{-\frac{1}{2}\sum_{\bf k}(\alpha_{{\bf k},0}-\alpha_{{\bf k},1})^2}\right]}.
\end{eqnarray}
At this point, the effective Hamiltonian of Eq.~(\ref{Hspinbosonnewbasis}) can be diagonalised and the ground state energy minimised as a function of $\alpha_{{\bf k},0}$ and $\alpha_{{\bf k},1}$ to find an optimal set of displacements in the ground state. For sub-Ohmic baths with $s<0.5$, this procedure has been shown to correctly capture the physics of the mean-field quantum phase transition of this model, and reproduces 
results obtained by non-perturbative numerical techniques.~\cite{chin11} However, for Ohmic systems we find that this procedure is unstable and deviates from the well-established results for the Ohmic case for $\alpha>1/2$. The cause for this is currently under investigation. Curiously, the correct behaviour is found for the unbiased case ($\epsilon=0$) by the variational transformation of Silbey and Harris,~\cite{silbey84} in which the constraint $\alpha_{{\bf k},0}=-\alpha_{{\bf k},1}$ is \emph{imposed}. However, as will be discussed further in Section~\ref{sec:SH}, the SH procedure fails for the biased case. 

We now propose a non-variational ansatz for the mode displacements for the biased Ohmic spin-boson model, appropriate for the scaling-limit ($\Delta/\omega_c\ll1$), which incorporates features of \emph{both} the sub-Ohmic and SH ground states, yet fixes the pathologies associated with both theories in this case. 
The proposed displacements are given by,~\cite{gan09}
\begin{eqnarray}
\label{ansatz0}
\alpha_{{\bf k},0}&{}={}&\frac{g_{\bf k}(\epsilon-\omega_{\bf k})}{\omega_{\bf k}(\omega_{\bf k}+\chi)},\label{alphadefn0}\\
\alpha_{{\bf k},1}&{}={}&\frac{g_{\bf k}(\epsilon+\omega_{\bf k})}{\omega_{\bf k}(\omega_{\bf k}+\chi)},\label{alphadefn1}
\label{ansatz1}
\end{eqnarray}
where $\chi=\sqrt{\Delta_r^2+\epsilon^2}$. This leads to a self-consistent equation for the renormalised tunneling given by
\begin{equation}\label{deltarselfcon}
\Delta_r=\Delta\exp{\left(-2\int_0^{\infty}d\omega\frac{J(\omega)}{(\omega+\chi)^2}\right)},
\end{equation}
which for $J(\omega)$ of Eq.~(\ref{spectraldensity}) becomes
%
\begin{equation}
\label{deltarOhmicSelf}
\Delta_r=\Delta\left(\frac{\chi}{\chi+\omega_c}\right)^{\alpha}e^{\alpha\omega_c/(\chi+\omega_c)}.
\end{equation}
%
Eq. (\ref{deltarselfcon}) self-consistently predicts the renormalised tunneling matrix element $\Delta_{r}$. It should be noted that in the presence of a bias, the integral is essentially cut off at low frequencies by the dynamical energy scale $\chi$, which is always non-zero for $\epsilon\neq0$. This means that in the biased case a finite solution to Eq.~(\ref{deltarOhmicSelf}) can always be found, and $\Delta_{r}$ is thus a \emph{continuous} function of $\alpha$. Similarly, we find (again for $J(\omega)$ of Eq.~(\ref{spectraldensity})) 
\begin{eqnarray}\label{A01ohmic}
A_0&=&\int_0^{\omega_c}d\omega\frac{J(\omega)(\epsilon-\omega)(2\chi+\epsilon+\omega)}{\omega(\omega+\chi)^2}\nonumber\\
&=&\frac{\alpha\omega_c}{2\chi(\chi+\omega_c)}(\epsilon^2+\chi(2\epsilon-\omega_c)),\\
A_1&=&\int_0^{\omega_c}d\omega\frac{J(\omega)(\epsilon+\omega)(\epsilon-\omega-2\chi)}{\omega(\omega+\chi)^2}\nonumber\\
&=&\frac{\alpha\omega_c}{2\chi(\chi+\omega_c)}(\epsilon^2-\chi(2\epsilon+\omega_c)),
\end{eqnarray}
which gives $R=A_0-A_1=2\alpha\omega_c\epsilon/(\chi+\omega_c)$. Hence, we may now write the ground state energy in the Ohmic case as
\begin{equation}\label{gsenergyohmic}
\lambda_0=\frac{1}{2}\left(\frac{\alpha\omega_c(\epsilon^2-\chi\omega_c)}{\chi(\chi+\omega_c)}-\eta\right),
\end{equation}
where $\eta=\sqrt{\Delta_r^2+\epsilon^2(1+2\alpha\omega_c/(\chi+\omega_c))^2}$, while the ground state magnetisation, $M=\langle \sigma_{z}\rangle$, and coherence, $\langle \sigma_{x}\rangle$, become
\begin{eqnarray}\label{Mohmic}
M=-\frac{\epsilon(1+2\alpha\omega_c/(\chi+\omega_c))}{\eta},
\end{eqnarray}
and
\begin{eqnarray}
\langle \sigma_{x}\rangle=\sqrt{1-M^2}\left(\frac{\Delta_{r}}{\Delta}\right).
\label{XOhmic}
\end{eqnarray}
The approximate ground state itself is written simply as
\begin{eqnarray}\label{gsansatz} 
\ket{\Psi_0}&=&-\frac{R+\epsilon-\eta}{\sqrt{\Delta_r^2+(R+\epsilon-\eta)^2}}|0\rangle\prod_{\bf k}D(\alpha_{{\bf k},0})|\rm{vac}\rangle\nonumber\\
&&+\frac{\Delta_r}{\sqrt{\Delta_r^2+(R+\epsilon-\eta)^2}}|1\rangle\prod_{\bf k}D(\alpha_{{\bf k},1})|\rm{vac}\rangle.\nonumber\\
\end{eqnarray}

\subsection{Comparison with the Silbey-Harris approach}
\label{sec:SH}

\begin{figure}
\begin{center}
\includegraphics[width=0.45\textwidth]{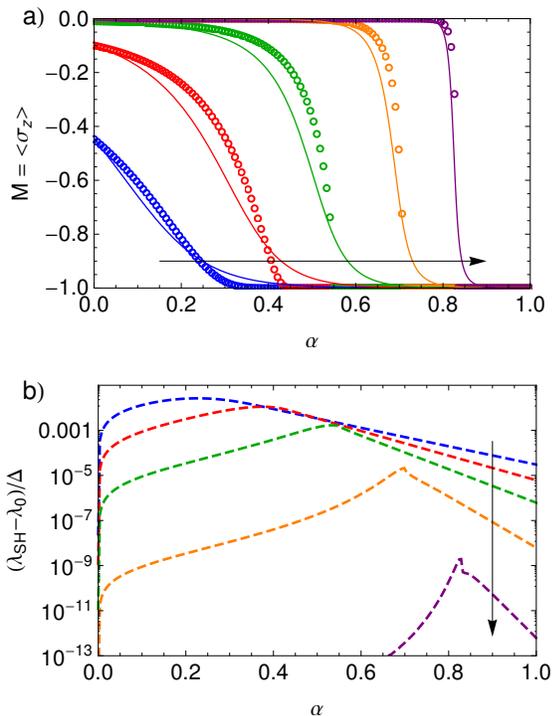}
\caption{(a) Magnetisation, $M$, as a function of system-bath coupling, $\alpha$, for the Ohmic spin-boson ground state. 
Results from the Silbey-Harris variational treatment are shown as open circles, while the solid curves are plotted using the ansatz presented in this work, see Eqs.~(\ref{Mohmic})~and~({\ref{gsansatz}}). Part (b) shows the difference in the ground state energies predicted by the Silbey-Harris theory and the present ansatz (both of which are negative). In both plots $\Delta=10^{-2}\omega_c$, with the arrows indicating values of $\epsilon=0.5\Delta$ (blue), $\epsilon=0.1\Delta$ (red), $\epsilon=10^{-2}\Delta$ (green), 
$\epsilon=10^{-4}\Delta$ (orange), and $\epsilon=10^{-8}\Delta$ (purple) in decreasing order.}
\label{SHcompM}
\end{center}
\end{figure}

We shall now compare our ground state ansatz [Eq.~(\ref{gsansatz})] 
to that given by the 
variational treatment of Silbey and Harris.~\cite{silbey84,harris85,silbey89,mccutcheon10} In the present formalism, the SH variational ground state is obtained by setting 
$\alpha_{{\bf k},0}=-\alpha_{{\bf k},1}=-g_{\bf k}/(\omega_{\bf k}+\Delta_{\mathrm{SH}}^2/\chi_{\mathrm{SH}})$ in Eqs.~(\ref{Hspinbosonnewbasis}) and~(\ref{deltardefn}), where 
$\Delta_{\mathrm{SH}}$ is the renormalised tunneling element found in the SH theory, and $\chi_{\mathrm{SH}}=\sqrt{\Delta_{\mathrm{SH}}^2+\epsilon^2}$. 
These displacements can be obtained by minimising the ground state energy of the Hamiltonian of Eq.~(\ref{Hspinbosonnewbasis}) with respect to 
$\alpha_{{\bf k},0}$ and $\alpha_{{\bf k},1}$, subject to the constraint $\alpha_{{\bf k},0}=-\alpha_{{\bf k},1}$ (which implies that $R\rightarrow0$). In doing so the self-consistent equation for 
$\Delta_{\mathrm{SH}}$ becomes,
\begin{equation}\label{deltarselfconSH}
\Delta_{\mathrm{SH}}=\Delta\exp{\left(-2\int_0^{\omega_c}\mathrm{d}\omega\frac{J(\omega)}{(\omega+\Delta_{\mathrm{SH}}^2/\chi_{\mathrm{SH}})^2}\right)},
\end{equation}
which leads to
\begin{equation}
\label{deltarselfconohmicSH}
\Delta_{\mathrm{SH}}=\Delta\left(\frac{\Delta_{\mathrm{SH}}^2}{\Delta_{\mathrm{SH}}^2+\chi_{\mathrm{SH}}\omega_c}\right)^{\alpha}\mathrm{exp}
\Big[\frac{\alpha\chi_{\mathrm{SH}}\omega_c}{\Delta_{\mathrm{SH}}^2+\chi_{\mathrm{SH}}\omega_c}\Big],
\end{equation}
again in the Ohmic case. 
Note that the low energy cut-off scale is now given by 
$\Delta_{\mathrm{SH}}^2/\chi_{\mathrm{SH}}$ which can self-consistently vanish above a critical coupling strength. 
This is the essence of the SH theory at strong coupling. The SH ground state energy is given by
\begin{equation}
\label{gsenergyohmicSH}
\lambda_{\mathrm{SH}}=\frac{1}{2}\left(\frac{\alpha\chi_{\mathrm{SH}}\omega_c^2}{(\Delta_{\mathrm{SH}}^2+\chi_{\mathrm{SH}}\omega_c)}-\chi_{\mathrm{SH}}\right),
\end{equation}
while we also find $M_{\mathrm{SH}}=\langle\sigma_z\rangle_{\rm{SH}}=-\epsilon/\chi_{\mathrm{SH}}$ and 
$\langle\sigma_x\rangle_{\mathrm{SH}}=\Delta_{\mathrm{SH}}^2/\chi_{\mathrm{SH}}\Delta$.

In Fig.~\ref{SHcompM} (a) we plot the magnetisation of the present spin-boson ground state ansatz (from Eq.~(\ref{Mohmic})) as a function of the dimensionless coupling $\alpha$, and compare with that given by the SH method, for various values of the bias $\epsilon$. As is now well known, for small $\epsilon$ the SH magnetisation, $M_{\mathrm{SH}}$, displays an unphysical discontinuous `jump' to $M_{\mathrm{SH}}=-1$ (corresponding to a fully localised ground state) for some value of $\alpha<1$. This behaviour can in turn be attributed to a \emph{discontinuous} change from non-zero $\Delta_{\mathrm{SH}}$ to $\Delta_{\mathrm{SH}}=0$ in the SH theory as $\alpha$ is varied. From Fig.~{\ref{SHcompM}} (a), however, we see that the ansatz presented in this work leads to no such `jumps' in the behaviour of $M$, and the magnetisation smoothly approaches $-1$ with increasing $\alpha$. 
This behaviour is in agreement with that found from various advanced numerical methods,~\cite{winter09, PhysRevLett98220401, Alvermann09,guo2011critical} as we shall show explicitly below using results derived from the Bethe-ansatz.

In Fig.~{\ref{SHcompM}} (b) we plot the difference in ground state energies predicted by the two theories, $\lambda_{\mathrm{SH}}-\lambda_0$, as a function 
of $\alpha$, for the same values of the bias as in part (a). We see that for all parameters considered here, the present ansatz corresponds to a ground state approximation 
with lower energy than that given by the SH theory (both $\lambda_{\mathrm{SH}}$ and $\lambda_0$ are negative for these parameters), suggesting that the state given by our ansatz is indeed a better approximation to the true ground state. Interestingly, we see that the difference in the two ground state energies is 
maximised through the crossover region, where $M_{\mathrm{SH}}$ is changing most rapidly. We also note that the unusual behaviour of $\lambda_{\mathrm{SH}}-\lambda_0$ near the peak difference is due to the discontinuities present in the values of $M_{\mathrm{SH}}$ and $\Delta_{\mathrm{SH}}$ as $\alpha$ is increased.

\subsection{Comparison with the Bethe-ansatz}


As discussed in Refs.~\onlinecite{leggett87,weissbook,PhysRevB324410,costi1998scaling,PhysRevB5912398,costi2003entanglement,PhysRevLett98220401,lehur08}, a mapping exists in the 
scaling limit ($\Delta/\omega_{c}\ll 1$) between the Ohmic spin-boson model, the 
anisotropic Kondo model, and a range of interacting resonance-level models. Exploiting this mapping and the existence of exact Bethe-ansatz solutions for the resonance-level 
model, explicit formulae for the properties of the biased spin-boson model can be obtained. To compare to the present ansatz results, we first take the Bethe-ansatz expressions for $\langle\sigma_z\rangle$ given in Eq.~(C1) of Ref.~\onlinecite{lehur08}. 

\begin{figure}
\begin{center}
\includegraphics[width=0.45\textwidth]{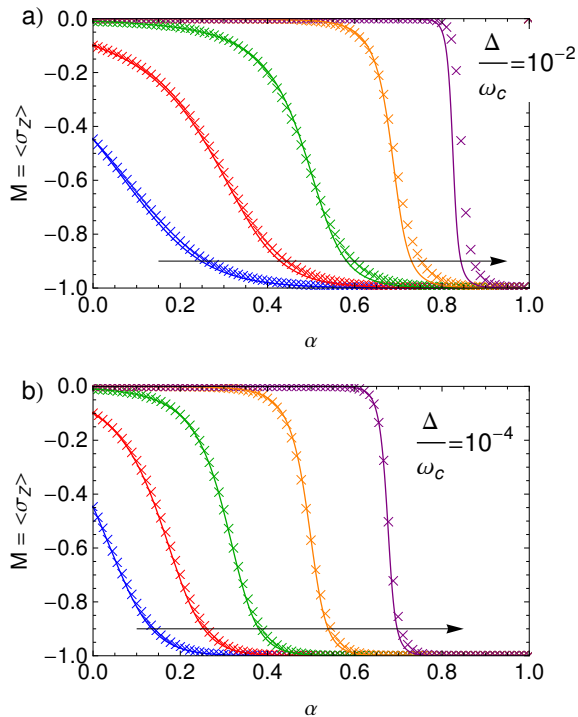}
\caption{Magnetisation, $M$, as a function of system-bath coupling, $\alpha$, calculated from the present ansatz (solid curves) and calculated using the exact scaling-limit expressions obtained via the Bethe ansatz (crosses), presented in Ref.~\onlinecite{lehur08}. The arrows indicate 
decreasing values of $\epsilon/\Delta$, as in Fig.~\ref{SHcompM}. The two plots correspond to different values 
of $\Delta/\omega_c$, as indicated.}
\label{BAcompM}
\end{center}
\end{figure}

\begin{figure}
\includegraphics[width=0.45\textwidth]{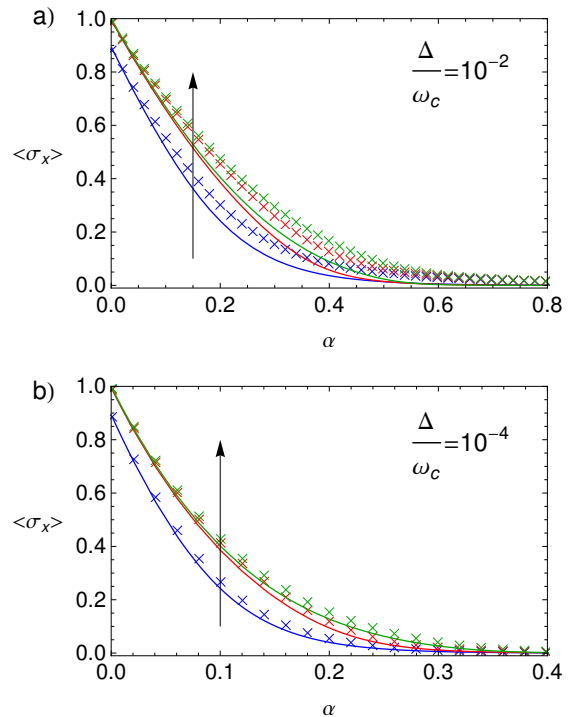}
\caption{Ground state spin coherence $\langle \sigma_{x}\rangle$ as a function of system-bath coupling, $\alpha$, calculated from the present ansatz (solid curves) and calculated using the exact scaling-limit expressions obtained via the Bethe ansatz (crosses), presented in Ref.~\onlinecite{lehur08}. The arrows indicate the bias values $\epsilon=0.5\Delta$ (blue), $\epsilon=0.1\Delta$ (red), and $\epsilon=10^{-2}\Delta$ (green), in decreasing order. 
The two plots correspond to different values of $\Delta/\omega_c$, as indicated.}
\label{sigmax}
\end{figure} 

In Fig.~\ref{BAcompM} we plot a comparison of the magnetisation predicted by our ansatz (solid curves), and that given by the Bethe-ansatz (crosses), for the same parameters as in Fig.~\ref{SHcompM}. We see that the agreement between the two methods is generally very good, though there are small deviations in the crossover behaviour at the weakest biases in plot (a). However, if we move further into the scaling-limit, by reducing the ratio $\Delta/\omega_c$ as in plot (b), we find excellent agreement across all bias values. This improvement with reducing ratio of $\Delta/\omega_c$ is not too surprising, as we expect both the present ground state and the Bethe-ansatz to be most valid in the scaling-limit. In the former case, due to the restricted basis from which the present ground state is constructed, in the latter case due to the mapping that is utilised from the spin-boson model to the anisotropic Kondo model, which is only valid in the limit that $\Delta/\omega_c$ becomes very small.~\cite{costi1998scaling,PhysRevB5912398,PhysRevB324410,lehur08} Still, it is remarkable how well the two methods agree, particularly over the sharp crossover region, given the simplicity of our proposed ground state [Eq.~(\ref{gsansatz})] in comparison to the full Bethe-ansatz.

Similar behaviour is seen for the ground state spin coherence $\langle \sigma_{x}\rangle$, 
as shown in Fig. \ref{sigmax}. 
Here, our ansatz leads to good qualitative agreement with the Bethe-ansatz but consistently predicts slightly lower coherences. Again, these deviations become smaller as we move further into the scaling limit. 
We note that the slightly weaker agreement for these quantities is likely to be due to the fact that they are non-universal and vanish in the scaling limit.~\cite{weissbook} The finite values we obtain thus depend on \emph{details} of the high frequency cut-off procedure. While the Bethe-ansatz results for the Kondo/resonance level model are still exact, the correspondence between them and the spin-boson model results is also dependent on the details of the high-frequency regularisation used in the mapping that links them in the scaling limit. As these details do not necessarily coincide for our ansatz and the Kondo mapping, numerical differences in coherences are likely to occur. However, as shown below, the leading-order functional forms of the non-universal properties are the same, and the numerical agreement for the magnetization (a universal property) is extremely good.   

\begin{figure}
\begin{center}
\includegraphics[width=0.45\textwidth]{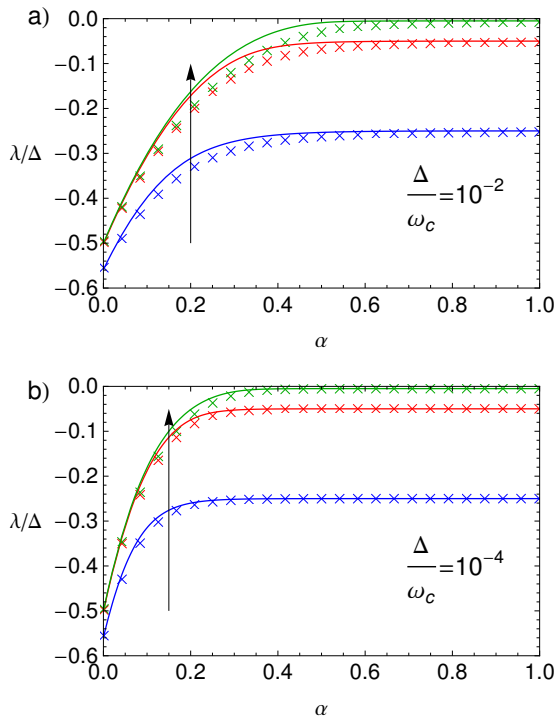}
\caption{Ground state energy, $\lambda$, as a function of system-bath coupling, $\alpha$, calculated from the present ansatz (solid curves) and calculated using expressions obtained via the Bethe-ansatz (crosses), presented in Ref.~\onlinecite{lehur08}. The arrows indicate the bias values $\epsilon=0.5\Delta$ (blue), $\epsilon=0.1\Delta$ (red), and $\epsilon=10^{-2}\Delta$ (green), in decreasing order. 
The two plots correspond to different values 
of $\Delta/\omega_c$, as indicated.}
\label{BAcompE}
\end{center}
\end{figure}

We can also compare the ground state energy predicted by our ansatz [Eq.~(\ref{gsenergyohmic})] with that given by the Bethe-ansatz (see Eqs.~(C3)-(C9) of Ref.~\onlinecite{lehur08}). 
Examples are shown in Fig.~\ref{BAcompE}, where we need to add a term equal to $\alpha\omega_c/2$ to Eq.~(\ref{gsenergyohmic}) to be consistent with the ground state energy definition in Refs.~\onlinecite{PhysRevB324410,leggett87,costi1998scaling,PhysRevB5912398,lehur08}. 
Once more, we see that the agreement improves as the ratio $\Delta/\omega_c$ decreases.

\subsubsection{Analytical results}\label{subsec:analytical}

Figs.~\ref{BAcompM},~\ref{sigmax}, and~\ref{BAcompE} suggest that, in the scaling-limit at least, the present ansatz provides a very good approximation to the true ground state of the Ohmic spin-boson Hamiltonian. We 
now investigate this further by comparing the analytic expressions obtained from this work with those presented in Ref.~\onlinecite{lehur08}. 
To recap, the magnetisation of the ground state in the present theory takes the form
$$M=-\frac{\epsilon(1+2\alpha\omega_{c}/(\chi+\omega_{c}))}{\eta},$$ where $\eta=\sqrt{\Delta_{r}^{2}+\epsilon^{2}(1+2\alpha\omega_{c}/(\chi+\omega_{c}))^{2}}$ and $\chi=\sqrt{\Delta_{r}^{2}+\epsilon^{2}}$. In the scaling limit, $\chi/\omega_{c}\rightarrow 0$, simple forms for the magnetisation and other parameters such as $\Delta_{r}$ can be obtained. 
When $\epsilon\ll\Delta_{r}$, we obtain to lowest order in $\epsilon/\omega_{c}$,
\begin{equation}M=-\frac{\epsilon(1+2\alpha)}{\Delta_{r}}\label{Mweak}.\end{equation}
The self-consistent equation for $\Delta_{r}$ in the scaling limit for weak bias is 
$$\Delta_{r}\approx\Delta e^{\alpha}\left(\frac{\Delta_{r}}{\omega_{c}}\right)^{\alpha},$$
which can be solved analytically and substituted into Eq. (\ref{Mweak}) to get,
\begin{eqnarray}
M&\propto&-\left(\frac{\epsilon}{T_{K}}\right),
\end{eqnarray}
where we have used the definition of the Kondo temperature, $T_{K}=\Delta\left({\Delta}/{D}\right)^{\alpha/1-\alpha}$, and the further relation between $D$ and $\omega_c$ 
given in Eq. (C8) of Ref. \onlinecite{lehur08} to facilitate an easier comparison with the Bethe-ansatz results. 
In this limit, we also obtain for the coherences, 
\begin{equation}
\langle \sigma_{x}\rangle \approx \frac{T_{K}}{\Delta}\left(1+\frac{\alpha}{2}\left(\frac{\epsilon}{T_{K}}\right)^{2}\right).
\end{equation}
For strong bias ($\epsilon\gg T_{K}$), we find that 
$$\Delta_{r}\approx\Delta\left(\frac{\epsilon}{\omega_{c}}\right)^\alpha,$$
and the magnetisation and coherence become
\begin{equation}
M+1\propto\left(\frac{\Delta}{\omega_{c}}\right)^{2}\left(\frac{\epsilon}{\omega_{c}}\right)^{2\alpha-2},
\end{equation}
and
\begin{equation}
\langle \sigma_{x}\rangle \propto\frac{\Delta}{\omega_{c}}\left(\frac{\epsilon}{\omega_{c}}\right)^{2\alpha-1},
\end{equation}
respectively, where, by our choice of notation, $M=-1$ is the magnetisation of the TLS in the absence of bath coupling in the limit $\epsilon/\Delta\gg1$. In both limits, these functional forms for the observables of the TLS coincide with the predictions of the Bethe-ansatz, and the cumbersome numerical prefactors we have omitted appear from our comparisons in Figs.~\ref{BAcompM},~\ref{sigmax}, and~\ref{BAcompE}
to be very close as well. 

\section{Unitary transformation approach}\label{unitary}

Having now shown that $\ket{\Psi_0}$ given in Eq.~(\ref{gsansatz}) can provide an excellent approximation to the ground state properties of the Ohmic spin-boson model, let's return to the original spin-boson Hamiltonian of Eq.~(\ref{Hspinboson}) to explore a unitary transformation approach to the problem. We shall show how this can be made equivalent to the ground state ansatz method outlined above, and how the transformation also provides a basis for computing the dynamics of the TLS beyond weak system-bath coupling. 

We perform the unitary transformation, $H'=e^SHe^{-S}$, where
\begin{equation}\label{exps}
e^{\pm S}=|0\rangle\langle0|\prod_{\bf k}D(\pm\delta_{{\bf k},0})+|1\rangle\langle1|\prod_{\bf k}D(\pm\delta_{{\bf k},1}),
\end{equation}
to give 
\begin{eqnarray}\label{Htransformed}
H'&=&\frac{\epsilon}{2}\sigma_z-\frac{\Delta_r}{2}\sigma_x+\sum_{\bf k}\omega_{\bf k}b_{\bf k}^{\dagger}b_{\bf k}-\frac{\Delta}{2}(\sigma_xB_x+\sigma_yB_y)\nonumber\\
&&+|0\rangle\langle0|(B_{z,0}+A_0')-|1\rangle\langle1|(B_{z,1}-A_1').
\end{eqnarray}
Here, $$B_x=\frac{B_++B_--2B}{2},\;\;\;B_y=\frac{B_--B_+}{2i},$$ $$B_{z,0}=\sum_{\bf k}(g_{\bf k}-\omega_{\bf k}\delta_{{\bf k},0})(b_{\bf k}^{\dagger}+b_{\bf k}),$$ and $$B_{z,1}=\sum_{\bf k}(g_{\bf k}+\omega_{\bf k}\delta_{{\bf k},1})(b_{\bf k}^{\dagger}+b_{\bf k}),$$ while $$A_{0}'=\sum_{\bf k}\delta_{{\bf k},0}(\omega_{\bf k}\delta_{{\bf k},0}-2g_{\bf k}),$$ $$A_{1}'=\sum_{\bf k}\delta_{{\bf k},1}(\omega_{\bf k}\delta_{{\bf k},1}+2g_{\bf k}),$$ and the renormalised tunneling is now given by $\Delta_r=B\Delta$, where we have defined $B={\rm tr}_B(B_{\pm}\rho_B)$, in terms of $B_{\pm}=\prod_{\bf k}D(\pm(\delta_{{\bf k},0}-\delta_{{\bf k},1}))$ and a thermal equilibrium bath state $\rho_B=\mathrm{exp}[-\beta\sum_{\bf k}\omega_{\bf k}b_{\bf k}^{\dagger}b_{\bf k}]/{\rm tr}_B(\mathrm{exp}[-\beta\sum_{\bf k}\omega_{\bf k}b_{\bf k}^{\dagger}b_{\bf k}])$. The inverse temperature is $\beta=1/k_BT$. 

Immediately, we see that if we choose $\delta_{{\bf k},0}=-\alpha_{{\bf k},0}$ and $\delta_{{\bf k},1}=-\alpha_{{\bf k},1}$, we get the same definitions of $A_0$ and $A_1$ as those used previously (i.e. $A_0'\rightarrow A_0$, $A_1'\rightarrow A_1$). Hence, we take
\begin{eqnarray}
\delta_{{\bf k},0}&{}={}&-\alpha_{{\bf k},0}=\frac{g_{\bf k}(\omega_{\bf k}-\epsilon)}{\omega_{\bf k}(\omega_{\bf k}+\chi)},\label{alphadefn0prime}\\
\delta_{{\bf k},1}&{}={}&-\alpha_{{\bf k},1}=-\frac{g_{\bf k}(\omega_{\bf k}+\epsilon)}{\omega_{\bf k}(\omega_{\bf k}+\chi)},\label{alphadefn1prime}
\end{eqnarray}
where $\chi=\sqrt{\Delta_r^2+\epsilon^2}$ as before. With these definitions, the renormalised tunneling becomes
\begin{eqnarray}\label{deltarfiniteT}
\Delta_r&{}={}&\Delta\exp{\left(-\frac{1}{2}\sum_{\bf k}(\delta_{{\bf k},0}-\delta_{{\bf k},1})^2\coth{\frac{\beta\omega_{\bf k}}{2}}\right)},\nonumber\\
&{}={}&\Delta\exp{\left(-2\int_0^{\omega_c}d\omega\frac{J(\omega)}{(\omega+\chi)^2}\coth{\frac{\beta\omega}{2}}\right)},
\end{eqnarray}
in the continuum limit, which is a finite temperature generalisation of Eq.~(\ref{deltarselfcon}).

Let us now split the transformed Hamiltonian as $H'=H_0'+H_I'$, where 
\begin{equation}\label{H0primed}
H_0'=\frac{\epsilon}{2}\sigma_z-\frac{\Delta_r}{2}\sigma_x+\sum_{\bf k}\omega_{\bf k}b_{\bf k}^{\dagger}b_{\bf k}+|0\rangle\langle0|A_0+|1\rangle\langle1|A_1,
\end{equation} and 
\begin{equation}\label{HIprimed}
H_I'=-\frac{\Delta}{2}(\sigma_xB_x+\sigma_yB_y)+|0\rangle\langle0|B_{z,0}-|1\rangle\langle1|B_{z,1},
\end{equation}
which ensures that $\langle H_I'\rangle_{H_0'}={\rm tr}[H_I'e^{-\beta H_0'}/{\rm tr}(e^{-\beta H_0'})]=0$. The Feynman-Bogolioubov upper bound on the free energy,~\cite{fisher} $A_B=-(1/\beta)\ln[{\rm tr}(e^{-\beta H_0'})]+\langle H_I'\rangle_{H_0'}+O(\langle H_I'^2\rangle_{H_0'})$, then becomes
\begin{equation}\label{AB}
A_B\approx\frac{A_0+A_1}{2}-\frac{1}{\beta}\ln{\left[2\cosh{\bigg(\frac{\beta\eta}{2}\bigg)}\right]},
\end{equation}
where $\eta=\sqrt{\Delta_r^2+(R+\epsilon)^2}$, with $R=A_0-A_1$, exactly as before, and we ignore contributions from the free bath Hamiltonian as we are interested only in the 
free energy of the TLS. In the zero temperature limit ($\beta\rightarrow\infty$) Eq.~({\ref{AB}}) becomes 
\begin{equation}\label{gsenergyAB}
A_B\rightarrow\lambda_0=\frac{1}{2}(A_0+A_1-\eta),
\end{equation}
in agreement with the ground state energy of Eq.~(\ref{gsenergyohmic}). 

Furthermore, approximating the thermal state density operator in the transformed frame as $\rho_{th}=e^{-\beta H_0'}/Z$, where $Z={\rm tr}(e^{-\beta H_0'})$, we find that the thermal expectation values of the Pauli spin operators may be written
\begin{equation}\label{thermalexpectation}
\langle\sigma_i\rangle={\rm tr}\left(e^S\sigma_ie^{-S}\rho_{th}\right).
\end{equation}
Hence, we find at finite temperature
\begin{equation}\label{Mdefnthermal}
\langle\sigma_z\rangle=-\frac{(R+\epsilon)}{\eta}\tanh{\bigg(\frac{\beta\eta}{2}\bigg)},
\end{equation}
and
\begin{equation}\label{sxdefnthermal}
\langle\sigma_x\rangle=\frac{\Delta_r^2}{\eta\Delta}\tanh{\bigg(\frac{\beta\eta}{2}\bigg)},
\end{equation}
both of which agree with our previous expressions (Eqs.~(\ref{Mohmic}) and~(\ref{XOhmic}), respectively) in the zero temperature limit.

\subsection{Dynamics - Master equation derivation}
\label{sec:Dynamics}

The advantage of the unitary transformation approach is that we are now in a position to efficiently explore the reduced TLS dynamics within the model, by deriving a master equation in the transformed representation. 
The general philosophy is that given we have shown that our zeroth-order Hamiltonian $H_0'$ provides a good approximation to the model ground state over a wide range of system-bath coupling strengths, we should expect (at low temperatures at least) that the effects of the perturbation $H_I'$ remain small over this parameter range too. Thus, we shall treat $H_I$ in low-order perturbation theory within the scaling limit, and derive a second-order master equation that should be valid at strong system-environment coupling as well as in the more usual weak-coupling regime. 

Using our partitioning of the total Hamiltonian given by Eqs.~({\ref{H0primed}}) and ({\ref{HIprimed}}), 
together with the time convolution-less projection operator technique,~\cite{b+p} we obtain an interaction 
picture time-local master equation of the form
\begin{eqnarray}\label{BM}
\frac{\partial{\tilde{\rho}}_v(t)}{\partial t}=-\int_0^{t}\mathrm{d}\tau{\rm tr}_B\left([H_I'(t),[H_I'(t-\tau),\tilde{\rho}_v(t)\rho_B]]\right),\nonumber\\
\end{eqnarray}
%
valid to second order in $H_I'$. Here, $\tilde{\rho}_v(t)$ is the reduced density operator of the TLS degrees of freedom in the transformed frame interaction picture, 
and the interaction 
Hamiltonian is $H'_I(t)=e^{iH'_0t}H'_Ie^{-iH'_0t}=-(\Delta/2)(\sigma_x(t)B_x(t)+\sigma_y(t)B_y(t))+\sigma_z(t)B_z(t)+\openone B_{I}(t)$, where we have defined the bath operators $B_z=(1/2)(B_{z,0}+B_{z,1})$ and $B_I=(1/2)(B_{z,0}-B_{z,1})$. We write $\sigma_i(t)=e^{iH_S't}\sigma_ie^{-iH_S't}$ and $B_i(t)=e^{iH_Bt}B_ie^{-iH_Bt}$, for $i=I,x,y,z$ (where $\sigma_I=\openone$), with $H'_S=((\epsilon+R)/2)\sigma_z-(\Delta_r/2)\sigma_x$ and $H_B=\sum_{\bf k}\omega_{\bf k}b_{\bf k}^{\dagger}b_{\bf k}$.
In deriving Eq.~({\ref{BM}}), both the reference state in the projection operator and the environment initial state (in the transformed frame) are chosen to be the thermal equilibrium 
state $\rho_B$, 
resulting in the absence of any inhomogeneous terms.~\cite{b+p,mccutcheon11_3,jang08,jang09}


We now insert the interaction Hamiltonian into Eq.~(\ref{BM}), take the trace over the environment, and make a secular approximation to remove fast oscillating terms.~\cite{b+p} This last simplification is made in order to ensure that our system relaxes to its ground state in the long time limit (at zero temperature), which is appropriate given the close agreement we find between the properties of our ansatz ground state and those of the Bethe ansatz.
Moving back to the Schr\"odinger picture, we then find 
\begin{align}
\label{TCL2homsimplifiedSchr}
\dot{\rho}_v=&-\frac{i}{2}[(\epsilon+R)\sigma_z-\Delta_r\sigma_x,{\rho}_v]\nonumber\\
&-\frac{\Delta^2}{4}\sum_{\lambda}\big\{[A_{x,\lambda},A_{x,\lambda}^{\dagger}{\rho}_{v}]\Gamma_{xx}(\lambda,t)\nonumber\\
&\;\;\;\;\;\;\;\;\;\;\;\;\;\;\;\;+[A_{y,\lambda},A_{y,\lambda}^{\dagger}{\rho}_{v}]\Gamma_{yy}(\lambda,t)+{\rm H.c.}\big\}\nonumber\\
&-\frac{\Delta}{2}\sum_{\lambda}\big\{\big([A_{y,\lambda},A_{z,\lambda}^{\dagger}{\rho}_{v}]
-[A_{z,\lambda},A_{y,\lambda}^{\dagger}{\rho}_{v}]\big)\Gamma_{yz}(\lambda,t)\nonumber\\
&\;\;\;\;\;\;\;\;\;\;\;\;\;\;\;\;+\delta_{\lambda,0}[A_{y,\lambda},\rho_{v}]\Gamma_{yI}(\lambda,t)+{\rm H.c.}\big\}\nonumber\\
&-\sum_{\lambda}\big\{[A_{z,\lambda},A_{z,\lambda}^{\dagger}{\rho}_{v}]\Gamma_{zz}(\lambda,t)\nonumber\\
&\;\;\;\;\;\;\;\;\;\;\;\;\;\;\;\;+\delta_{\lambda,0}[A_{z,\lambda},\rho_{v}]\Gamma_{zI}(\lambda,t)+{\rm H.c.}\big\},
\end{align}
%
%
where $\delta_{\lambda,0}$ is the Kronecker delta, and we have decomposed the system operators as $\sigma_i(t)=\sum_{\lambda}e^{i\lambda t}A_{i,\lambda}$, for $\lambda=0,\pm\eta$. In terms of $\theta=\arctan(\frac{\Delta_r}{\epsilon+R})$, we have $A_{x,0}=\sin\theta(\sin\theta\sigma_x+\cos\theta\sigma_z)$,  $A_{x,\eta}=\frac{\cos\theta}{2}(\cos\theta\sigma_x+i\sigma_y-\sin\theta\sigma_z)$,  $A_{y,0}=0$,  $A_{y,\eta}=-\frac{i}{2}(\cos\theta\sigma_x+i\sigma_y-\sin\theta\sigma_z)$,  $A_{z,0}=\cos{\theta}(\sin\theta\sigma_x+\cos\theta\sigma_z)$,  $A_{z,\eta}=\frac{\sin\theta}{2}(-\cos\theta\sigma_x-i\sigma_y+\sin\theta\sigma_z)$, while $A_{i,-\eta}=A_{i,\eta}^{\dagger}$. The bath response functions 
\begin{equation}\label{bathresponse}
\Gamma_{ij}(\lambda,t)=\int_0^{t}d\tau e^{i\lambda\tau}\Lambda_{ij}(\tau),
\end{equation}
are defined in terms of the 
correlation functions 
$\Lambda_{ij}(\tau)={\rm tr}_B(B_i(\tau)B_j(0)\rho_B)$, for $i,j=I,x,y,z$. 

We proceed by writing 
the transformation parameters as
\begin{eqnarray}
\delta_{{\bf k},0}&{}={}&\zeta_{{\bf k},0}\frac{g_{\bf k}}{\omega_{\bf k}},\label{zeta0defn}\\
\delta_{{\bf k},1}&{}={}&\zeta_{{\bf k},1}\frac{g_{\bf k}}{\omega_{\bf k}},\label{zeta1defn}
\end{eqnarray}
such that $\zeta_{{\bf k},0}=(\omega_{\bf k}-\epsilon)/(\omega_{\bf k}+\chi)$ and $\zeta_{{\bf k},1}=-(\omega_{\bf k}+\epsilon)/(\omega_{\bf k}+\chi)$ 
for our ansatz [see Eqs.~(\ref{alphadefn0prime}) and~(\ref{alphadefn1prime})]. We may now re-write the renormalised tunneling as
\begin{eqnarray}\label{deltarfiniteTzeta}
\Delta_r&=&\Delta\exp{\left(-\frac{1}{2}\sum_{\bf k}\frac{g_{\bf k}^2}{\omega_{\bf k}^2}(\zeta_{{\bf k},0}-\zeta_{{\bf k},1})^2\coth{\frac{\beta\omega_{\bf k}}{2}}\right)},\nonumber\\
&=&\Delta\exp{\left(-\frac{1}{2}\int_0^{\omega_c}d\omega\frac{J(\omega)}{\omega^2}\zeta_-(\omega)^2\coth{\frac{\beta\omega}{2}}\right)},\nonumber\\
\end{eqnarray}
where $\zeta_{\pm}(\omega)=\zeta_{0}(\w)\pm\zeta_{1}(\w)$, with $\zeta_{0}(\w)=(\omega-\epsilon)/(\omega+\chi)$ and $\zeta_{1}(\w)=-(\omega+\epsilon)/(\omega+\chi)$ in the continuum limit.  
In terms of $\zeta_{\pm}(\omega)$ 
the correlation functions are found to read
\begin{eqnarray}
\Lambda_{xx}(\tau)&{}={}&\frac{1}{2}(C(\tau)+G(\tau)-2B^2),\\
\Lambda_{yy}(\tau)&{}={}&\frac{1}{2}(C(\tau)-G(\tau)),
\end{eqnarray}
where
\begin{eqnarray}
C(\tau)&=&{\rm exp}\bigg[-\int_0^{\infty}d\omega\frac{J(\omega)}{\omega^2}\zeta_-(\omega)^2\nonumber\\
&&\;\;\;\;\;\;\;\;\;\;\;\;\;\;\;\;\times\left((1-\cos{\omega\tau})\coth\beta\omega/2+i\sin{\omega\tau}\right)\bigg],\nonumber\\
G(\tau)&=&{\rm exp}\bigg[-\int_0^{\infty}d\omega\frac{J(\omega)}{\omega^2}\zeta_-(\omega)^2\nonumber\\
&&\;\;\;\;\;\;\;\;\;\;\;\;\;\;\;\;\times((1+\cos{\omega\tau})\coth\beta\omega/2-i\sin{\omega\tau})\bigg],\nonumber\\
\end{eqnarray}
and $B=\Delta_r/\Delta$, while
\begin{eqnarray}
\Lambda_{yz}(\tau)&=&-B\int_0^{\infty}d\omega\frac{J(\omega)}{\omega}\zeta_-(\omega)\left(1-\frac{\zeta_-(\omega)}{2}\right)\nonumber\\
&&\;\;\;\;\;\;\;\;\;\;\;\;\;\times(\sin{\omega\tau}\coth\beta\omega/2+i\cos{\omega\tau}),\nonumber\\
\Lambda_{yI}(\tau)&=&+B\int_0^{\infty}d\omega\frac{J(\omega)}{\omega}\frac{\zeta_-(\omega)\zeta_+(\omega)}{2}\nonumber\\
&&\;\;\;\;\;\;\;\;\;\;\;\;\;\times(\sin{\omega\tau}\coth\beta\omega/2+i\cos{\omega\tau}),\nonumber\\
\Lambda_{zz}(\tau)&=&\int_0^{\infty}d\omega J(\omega)\left(1-\frac{\zeta_-(\omega)}{2}\right)^2\nonumber\\
&&\;\;\;\;\;\;\;\;\;\;\;\;\;\times(\cos{\omega\tau}\coth\beta\omega/2-i\sin{\omega\tau}),\nonumber\\
\Lambda_{zI}(\tau)&=&-\int_0^{\infty}d\omega J(\omega)\frac{\zeta_+(\omega)}{2}\left(1-\frac{\zeta_-(\omega)}{2}\right)\nonumber\\
&&\;\;\;\;\;\;\;\;\;\;\;\;\;\times(\cos{\omega\tau}\coth\beta\omega/2-i\sin{\omega\tau}),\nonumber\\
\end{eqnarray}
and we have already used the fact that 
$\Lambda_{xy}(\tau)=\Lambda_{yx}(\tau)=\Lambda_{xz}(\tau)=\Lambda_{zx}(\tau)=\Lambda_{xI}(\tau)=\Lambda_{Ix}(\tau)=0$ in deriving Eq.~({\ref{TCL2homsimplifiedSchr}}). We
note that Eq.~({\ref{TCL2homsimplifiedSchr}}) and the subsequent definitions are quite general, and the (secular) master equation corresponding to any polaron-like transformation on the 
Hamiltonian can be obtained by the appropriate choice of the functions $\zeta_{0}(\w)$ and $\zeta_{1}(\w)$. For example, the master 
equation corresponding to the SH transformation~\cite{mccutcheon11_2,mccutcheon11_3} is given by setting 
$\zeta_0^{SH}(\omega)=\omega/(\omega+\Delta_{\mathrm{SH}}/\chi_{\mathrm{SH}})$ and $\zeta_1^{SH}(\omega)=-\omega/(\omega+\Delta_{\mathrm{SH}}/\chi_{\mathrm{SH}})$, which we shall use for comparison to our ansatz below.

We now solve the ansatz master equation numerically to determine the dissipative spin dynamics, concentrating on the zero temperature limit. At zero temperature, and for 
an Ohmic spectral density, the correlation functions defined above do not decay on a rapid timescale compared to the system dynamics we wish to capture. 
As such, in the following we cannot extend the integration limits in Eq.~({\ref{bathresponse}}) to infinity. We consider the system to be initialised in state $\ket{0}$ at time $t=0$.

\begin{figure}
\includegraphics[width=0.45\textwidth]{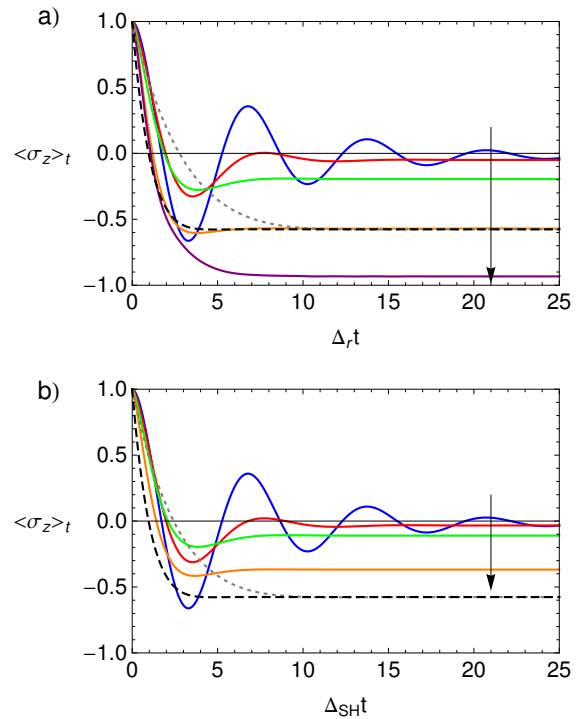}
 \caption{Magentisation dynamics $\langle \sigma_{z}\rangle_t$ computed within the secular approximation as a function of (a) $\Delta_{r}t$ for the present theory and (b) $\Delta_{SH}t$ for Silbey-Harris theory. Solid curves: the arrows indicate the coupling strengths $\alpha=0.1$ (blue), $\alpha=0.25$ (red), $\alpha=0.4$ (green), $\alpha=0.5$ (orange), and $\alpha=0.6$ (purple, shown for our ansatz only) in increasing order. The dotted curves correspond to the exact (scaling limit) solution at $\alpha=0.5$, extracted from a mapping to the Toulouse Hamiltonian, and computed from the formula given in Ref.~\onlinecite{weissbook}. The dashed curves show the same exact solution, but now plotted in units of $T_{k}t$, where $T_{k}=\Delta^2/\omega_{c}$ is the Kondo scale for the Toulouse problem.~\cite{weissbook} Other parameters: $\Delta=10^{-2}\w_c$, $\epsilon=10^{-2}\Delta$, and $T=0$.}
\label{secular}
\end{figure}

In Fig.~{\ref{secular}} we plot the population dynamics, $\langle\sigma_z\rangle_t={\rm tr}(\sigma_z\rho_v(t))$, using (a) our present ansatz, and (b) the SH theory, as a function of the relevant scaled time, i.e.  $\Delta_rt$ and $\Delta_{\rm SH}t$, respectively. The solid curves in (a) show coupling strengths ranging from $\alpha=0.1$ to $\alpha=0.6$ for the present ansatz, while in (b) the range $\alpha=0.1$ to $\alpha=0.5$ is plotted using the SH theory (in this case the $\alpha=0.6$ curve is not plotted as $\Delta_{\rm SH}=0$ already for the SH theory). The other parameters are 
$\Delta=10^{-2}\w_c$, $\epsilon=10^{-2}\Delta$, and $T=0$, which correspond to the central green curve in Fig.~{\ref{SHcompM}}.
For the smallest coupling strength ($\alpha=0.1$), 
we see that the dynamics calculated from the present ansatz and from the SH transformation agree almost perfectly. This is  unsurprising, as we know from our previous comparisons that for these parameters the ground states given by the two methods are very similar. Nevertheless, discrepancies do begin to become apparent between the two methods at larger $\alpha$, as shown by the curves corresponding to $\alpha=0.25$, $\alpha=0.4$, and $\alpha=0.5$. In particular, we see that in these cases the ansatz steady-states settle at long times to slightly lower values of $\langle\sigma_z\rangle$ than the SH dynamics predicts. This is to be expected, given the corresponding ground state magnetisation behaviour shown in Fig.~\ref{SHcompM}, and the fact that the two-state system relaxes towards the relevant ground state at long times for $T=0$ in the secular master equation.
At coupling strengths beyond $\alpha\sim0.55$, the SH theory predicts a complete renormalisation of the tunneling strength to zero, i.e. $\Delta_{\mathrm{SH}}\rightarrow0$, while for $\alpha<1$ the tunneling always remains non-zero when using the present ansatz. Hence, significant differences emerge in the population dynamics, 
with the two approaches disagreeing on both the population decay rates and steady-states. In particular, as a consequence of $\Delta_{\mathrm{SH}}\rightarrow0$ for $\alpha>0.55$, 
the SH theory always incorrectly predicts a system steady-state that is completely localised in this regime, which is not the case for our ansatz. For biased systems at $T=0$, this failing is also well-known to be true of the non-interacting blip approximation (for all $\alpha<1$).~\cite{leggett87,weissbook} Again, the problem can be traced back to the incorrect choice of zeroth-order Hamiltonian in the SH theory (and in effect within the non-interacting blip approximation as well). 


\subsubsection{The Toulouse point}
\label{subsec:toulouse}
We have shown that our unitary transformation approach 
(correctly) leads to significant changes in the spin dynamics at strong coupling compared to the SH theory. However, to assess the 
accuracy of the new dynamics, we now compare these results 
to the exact expressions 
that are known at the 
Toulouse point of the spin-boson model, which occurs at $\alpha=0.5$. At this coupling, the spin-boson model becomes equivalent to a non-interacting resonance level model \emph{in the scaling limit}, and the spin dynamics can be solved exactly.~\cite{leggett87, weissbook}


The dotted curves in Fig.~\ref{secular} show the exact analytical solution for $\alpha=0.5$ taken from Ref.~\onlinecite{weissbook}, as a funnction of scaled time $\Delta_rt$ in (a) and $\Delta_{\rm SH}t$ in (b).
We see that the final value of $\langle \sigma_{z}\rangle$ within our ansatz coincides with that of the exact Toulouse point solution in (a), while, as expected, the SH theory dynamics in (b) tends to a substantially less magnetised, and incorrect, final state. We observe, however, that the dynamics of the Toulouse point solution appears to be slower than that predicted by our master equation approach. We suggest that the origin of these different decay rates could be related to the non-universal factors which appear in the ansatz predictions for $\Delta_{r}$.  Specifically, 
the exponential factor that appears in the self-consistent equation for $\Delta_{r}$ [Eq. (\ref{deltarOhmicSelf})] is not a universal feature, in the sense that it depends on the way that integrals over the bath spectral function are cut-off for $\omega>\omega_{c}$. For instance, using an exponential cut-off in the Ohmic spectral function, $ J(\omega)=(\alpha/2)\omega e^{-\omega/\omega_{c}}$, yields the following self-consistent equation for $\Delta_{r}$ in the scaling limit: 
\begin{equation}
\Delta_{r}=\Delta\left(\frac{\chi}{\omega_{c}}\right)^{\alpha}e^{\alpha(1+\gamma)},
\end{equation}
where $\gamma\approx 0.577$ is the Euler-Mascheroni constant, as compared to $\Delta_{r}=\Delta\left(\chi/\omega_{c}\right)^{\alpha}e^{\alpha}$ for $J(\omega)=(\alpha/2)\omega\theta(\omega_{c}-\omega)$. In the absence of a bias, we see that the solution of the self-consistent equations then leads to an effective Kondo temperature $T_{K}=\Delta_{r}=\Delta\left(\Delta f/\omega_{c}\right)^{\alpha/(1-\alpha)}$, where $f$ is numerical factor ($\approx e$) which depends on the form of the cut-off. The factor $f$ is non-universal, and this is also true of SH theory. In the exact solution presented 
in Ref.~\onlinecite{weissbook} for the Toulouse point, the Kondo temperature that appears has $f=1$, and is therefore smaller than the $T_{k}$ predicted by our ansatz or SH theory. Therefore, when we plot 
dynamics in units of $\Delta_{r}t$ or $\Delta_{\rm SH}t$, 
computed from Eqs.~(\ref{deltarOhmicSelf}) and~(\ref{deltarselfconohmicSH}), respectively, we are not using the natural units for the Toulouse point solution. 

The dashed curves in Fig.~\ref{secular} show 
the Toulouse point solution when 
plotted as a function of $\Delta^2 t/\omega_{c}$, which is the appropriate Kondo temperature at $\alpha=0.5$. The ansatz and Toulouse point curves in (a) now show much closer agreement, whereas SH theory still captures neither the right dynamical timescale nor, of course, the final magnetisation. 


These results suggest more generally 
that variational or 
ansatz-based polaron 
methods of the kind explored here may 
suffer from artefacts arising from the treatment of low frequency modes, which lead to dependencies of the effective Hamiltonian parameters on the details of the high frequency cut-off procedure. As far as we are aware, this has not previously been discussed or explored in detail, and is usually neglected. We have seen here, however, that this may have a material effect on the dynamics of the system, and would therefore be an interesting issue to explore in future work. 

\section{Non-Ohmic spectral densities}
\label{sec:other}
In this section we briefly comment on the application of our generalised polaron theory to non-Ohmic environments. However, for reasons to be set out below, we believe that the particular form of our ansatz is likely only to be useful in the Ohmic case.  Although the sudden collapse of coherent tunnelling is not an issue with super-Ohmic baths, we could expect that our ansatz/unitary transformation may still lead to different dynamics in such a case 
compared to the SH theory, for example through the change in dynamical cut-off in the renormalised tunnelling matrix element and the bath-induced bias (which does not feature in SH theory at all). However, repeating the calculations of Section~\ref{sec:ansatz} 
for spectral densities of the form $J(\omega)\propto \omega^s$ (with $s>1$) shows that the effects of asymmetric displacements in the renormalised tunnelling amplitude and bath-induced magnetisation vanish in the scaling limit. As renormalisation effects converge to those predicted by adiabatic renormalisation, or, equivalently, standard polaron theory, (as they also do in SH theory) we therefore expect this theory to coincide with full polaron theory for super-Ohmic baths as $\omega_{c}\rightarrow\infty$.   

As we have previously mentioned, the asymmetric displacement terms are extremely important in the sub-Ohmic case, and may appear spontaneously even without a bias at strong coupling and $T=0$.~\cite{chin11} Repeating the calculations of Section~\ref{sec:ansatz} shows this time that the effective bias induced by the environment is proportional to $({\omega_{c}}/{\chi})^{1-s}$, and thus diverges in the scaling limit. This would predict complete localisation for any alpha and would be inconsistent with the known existence of a quantum phase transition in sub-Ohmic systems.~\cite{spohn85,kehrein, vojta2005quantum, winter09,Alvermann09,LeHur07, chin11, Zhang10,florens2011dissipative,zhao11} However, the variational ansatz proposed by Chin et al. in Ref.~\onlinecite{chin11} regulates this pathology, 
and could also be used as the basis for a variational unitary transformation that is similar to the dynamical approach pursued here. 
This will be discussed in a forthcoming work. 
 
\section{Conclusions}
\label{sec:Conclusions}
We have presented a new ansatz for the ground state of the biased Ohmic spin-boson model which is of lower energy than the Silbey-Harris state and cures the problem of the discontinuous collapse of the coherent tunneling matrix element as the coupling strength to the environment is increased. The key differences between our ansatz and the SH variational state result from the different forms of the displacements $\alpha_{{\bf k},0}$ and $\alpha_{{\bf k},1}$ which are taken in these ground states. Firstly, the SH ground state has strictly (anti) symmetric displacements and the bias only appears in the low frequency energy scale $\Delta_{\mathrm{SH}}^{2}/\chi_{\mathrm{SH}}$. In our ansatz the displacements are asymmetric due to the inclusion of the bias in the numerators of Eqs. (\ref{alphadefn0}) and (\ref{alphadefn1}). Similar asymmetric displacements were shown to be essential for describing the spontaneous magnetization of the unbiased sub-Ohmic spin-boson model which characterises its quantum phase transition, and in effect causes an extra bath-induced bias to be seen by the TLS.~\cite{chin11,zhao11} Moreover, we have shown explicitly that asymmetric displacements lead to the appearance of the correct dynamical low energy cut-off ($\chi$) in the self-consistent equation for $\Delta_{r}$ for the biased Ohmic case. The close agreement of results obtained through our ansatz and the Bethe-ansatz evidences that our ansatz is capturing a considerable amount of the essential physics of the problem. However, the detailed links between the rather simple form of the spin-boson model ground state we have given and that of the equivalent Kondo model remain to be analyzed, particularly for non-universal observables, such as the spin coherence, and outside of the scaling limit. 


Importantly, by showing that this ground state can be obtained via a unitary transformation technique, it is now possible to treat the dynamics of the Ohmic spin-boson model using master equation techniques from polaron theory. The examples given in this article already illustrate the dramatic differences in relaxation behaviour, particularly in terms of steady states, 
when the zeroth-order Hamiltonian correctly preserves a coherent tunneling matrix elements at all coupling strengths. This may be of relevance for parameter regimes found in some biomolecular complexes, where biases, environmental coupling strengths and (bare) tunneling amplitudes are comparable,~\cite{ishizaki2010quantum} conditions under which the differences between SH theory and the present theory are greatest (at $T=0$). Although some deviation was found between our results and the nominally exact dynamics at the Toulouse point, it is clear that our theory still performs better than SH theory at this point, and we were able to understand where these differences arise. While our ansatz cures one of the most important problems associated with SH theory in biased Ohmic systems, insights such as those given above also point the way to future refinements of the present theory. 

Finally, our ansatz could also be applied using the more advanced, non-Markovian master equation techniques, such as those presented and analysed in Refs.~\onlinecite{jang08, jang09, mccutcheon11_3,cheekong12}, which can also include bath relaxation effects. These improvements in the handling of the dynamics could go some way to reducing some of the differences we found with the Toulouse solution, and along with effects due to finite temperatures and moving beyond the scaling limit, are interesting areas for investigation in the near future.

\section*{Acknowledgements}

AN would like to thank Imperial College for financial support, DPSM acknowledges support from CHIST-ERA project SSQN and the EPSRC, and AWC is supported by The Winton Programme for the Physics of Sustainability.

\end{document}